\documentclass[letter]{aa} 

\usepackage{graphicx,url,twoopt,natbib}
\usepackage[varg]{txfonts}
\usepackage[pdftitle={The formation of the Galilean moons and Titan in the Grand Tack scenario}, colorlinks = true, breaklinks = true, citecolor = blue, linkcolor = blue, urlcolor = blue, pdfauthor = {Ren\'{e} Heller}]{hyperref}
\usepackage{breakurl}
\bibpunct{(}{)}{;}{a}{}{,}    

\newcommand\KG[1]{#1}      
\newcommand\stumm[1]{}      

\begin{document} 

\title{The formation of the Galilean moons and Titan \\ in the Grand Tack scenario}

\author{R.~Heller\inst{1,2}\fnmsep\thanks{Postdoctoral Fellow, Canadian Astrobiology Training Program}
           \and
           G.\KG{-}D.~Marleau\inst{3}\,\fnmsep\thanks{\KG{Member of the International Max Planck Research School}}
           \and
           R.~E.~Pudritz\inst{1,2}
           }

\institute{Origins Institute, McMaster University, 1280 Main Street West, Hamilton, ON L8S 4M1, Canada
              \and
              Department of Physics and Astronomy, McMaster University, \href{mailto:rheller@physics.mcmaster.ca}{rheller@physics.mcmaster.ca} | \href{mailto:pudritz@physics.mcmaster.ca}{pudritz@physics.mcmaster.ca}
              \and
              Max-Planck-Institut f\"ur Astronomie, K\"onigstuhl 17, 69117 Heidelberg, Germany, \href{mailto:marleau@mpia.de}{marleau@mpia.de}
              }

\date{Received 17 April 17 2015 / Accepted 2 June 2015}
 
\abstract
{In the Grand Tack (GT) scenario for the young solar system, Jupiter formed beyond 3.5\,AU from the Sun and migrated as close as 1.5\,AU until it encountered an orbital resonance with Saturn. Both planets then supposedly migrated outward for several $10^5$\,yr, with Jupiter ending up at $\approx5$\,AU. The initial conditions of the GT and the timing between Jupiter's migration and the formation of the Galilean satellites remain unexplored.}
{We study the formation of Ganymede and Callisto, both of which consist of $\approx50$\,\% H$_2$O and rock, in the GT scenario. We examine why they lack dense atmospheres, while Titan is surrounded by a thick N$_2$ envelope.}
{We model an axially symmetric circumplanetary disk (CPD) in hydrostatic equilibrium around Jupiter. The CPD is warmed by viscous heating, Jupiter's luminosity, accretional heating, and the Sun. The position of the H$_2$O ice line in the CPD, which is crucial for the formation of massive moons, is computed at various solar distances. We assess the loss of Galilean atmospheres due to high-energy radiation from the young Sun.}
{Ganymede and Callisto cannot have accreted their H$_2$O during Jupiter's supposed GT, because its CPD (if still active) was too warm to host ices and much smaller than Ganymede's contemporary orbit. From a thermal perspective, the Galilean moons might have had significant atmospheres, but these would probably have been eroded during the GT in $<10^5$\,yr by solar XUV radiation.}
{Jupiter and the Galilean moons formed beyond $4.5\,\pm\,0.5$\,AU and prior to the proposed GT. Thereafter, Jupiter's CPD would have been dry, and delayed accretion of planetesimals should have created water-rich Io and Europa. While Galilean atmospheres would have been lost during the GT, Titan would have formed after Saturn's own tack, because Saturn still accreted substantially for $\approx10^6$\,yr after its closest solar approach, ending up at about 7\,AU.}


\maketitle

\section{Introduction}

Recent simulations of the early solar system suggest that the four giant planets underwent at least two epochs of rapid orbital evolution. In the Grand Tack (GT) model \citep{2011Natur.475..206W}, they formed beyond the solar water (H$_2$O) ice line \citep{2006Icar..181..178C} between 3.5\,AU and 8\,AU. During the first few $10^6$\,yr in the protostellar disk, when Jupiter had fully formed and Saturn was still growing, Jupiter migrated as close as 1.5\,AU to the Sun within some $10^5$\,yr, until Saturn gained enough mass to migrate even more rapidly, catching Jupiter in a 3:2 or a 2:1 mean motion resonance \citep{2014ApJ...795L..11P}. Both planets then reversed their migration, with Jupiter ending up at approximately 5\,AU. The other gas planets also underwent rapid orbital evolution due to gravitational interaction, pushing Uranus and Neptune to and beyond about 10\,AU, respectively, while Saturn settled at $\approx7$\,AU. A second period of rapid orbital evolution occurred $\approx7\times10^8$\,yr later, according to the Nice model \citep{2005Natur.435..459T,2005Natur.435..462M,2005Natur.435..466G}, when Jupiter and Saturn crossed a 2:1 mean motion resonance, thereby rearranging the architecture of the gas giants and of the minor bodies.

While the GT delivers adequate initial conditions for the Nice model, the initial conditions of the GT itself are not well constrained \citep{2014arXiv1409.6340R}. Details of the migrations of Jupiter and Saturn depend on details of the solar accretion disk and planetary accretion, which are also poorly constrained \citep{2014RSPTA.372.0174J}. We note, however, that the moons of the giant planets provide additional constraints on the GT that have scarcely been explored. As an example, using $N$-body simulations, \citet{2011A&A...536A..57D} studied the orbital stability of the Uranian satellites during the Nice instability. Hypothetical moons beyond the outermost regular satellite Oberon typically got ejected, while the inner moons including Oberon remained bound to Uranus. Their results thus support the validity of the Nice model. Later, \citet{2014AJ....148...25D} found that one of three Jovian migration paths proposed earlier \citep{2012AJ....144..117N} is incompatible with the orbital stability and alignment of the Galilean moons, yielding new constraints on the Nice model.

To our knowledge, no study has used the Galilean moons or Saturn's major moon Titan, which is surrounded by a thick nitrogen (N$_2$) atmosphere, to test the plausibility of the GT model. In this Letter, we focus on the early history of the icy Galilean moons, Ganymede and Callisto, and on Titan in the GT scenario; assuming that the GT scenario actually took place in one form or another, we identify new constraints on the timing of their formation. The novel aspect of our study is the evolution of the H$_2$O ice line in Jupiter's circumplanetary disk (CPD) under the effect of changing solar illumination during the GT. As the position and evolution of the ice line depends on irradiation from both the forming Jupiter and the Sun \citep{2015A&A...578A..19H}, we want to understand how the observed dichotomy of two mostly rocky and two very icy Galilean moons can be produced in a GT setting.

\vspace{-.19cm}

\section{The icy Galilean moons in the Grand Tack model}

\subsection{Formation in the circumjovian accretion disk}
\label{sub:formation}

Both Ganymede and Callisto consist of about 50\,\% rock and 50\,\% H$_2$O, while the inner Galilean satellites Io and Europa are mostly rocky. This has been considered a record of the temperature distribution in Jupiter's CPD at the time these moons formed \citep{1974Icar...21..248P,2010ApJ...714.1052S}. In particular, the H$_2$O ice line, which is the radial distance at which the disk is cool enough for the transition of H$_2$O vapor into solid ice, should have been between the orbits of rocky Europa at about $9.7$ Jupiter radii ($R_{\rm Jup}$) and icy Ganymede at about $15.5\,R_{\rm Jup}$ from Jupiter.

We simulate a 2D axisymmetric CPD in hydrostatic equilibrium around Jupiter using a ``gas-starved'' standard disk model \citep{2002AJ....124.3404C,2006Natur.441..834C} that has been modified to include various heat sources \citep{2014SoSyR..48...62M}, namely, (i) viscous heating, (ii) planetary illumination in the ``cold-start scenario'', (iii) direct accretion onto the CPD, and (iv) stellar illumination. The model is coupled to pre-computed planet evolution tracks \citep{2013A&A...558A.113M}, and we here assume a solar luminosity 0.7 times its current value to take into account the faint young Sun \citep{1972Sci...177...52S}. Details of our semi-analytical model are described in \citet{2015A&A...578A..19H,2015ApJ...806..181H}.

We evaluate the radial position of the circumjovian H$_2$O ice line at different solar distances of Jupiter and study several disk opacities ($\kappa_{\rm P}$) as well as different shutdown accretion rates for moon formation ($\dot{M}_{\rm shut}$), all of which determine the radial distributions of the gas surface density and midplane temperature in the CPD. We follow the CPD evolution after the planet opens up a gap in the circumstellar disk (CSD) until the planetary accretion rate ($\dot{M}_{\rm p}$) in the pre-computed track drops to a particular value of $\dot{M}_{\rm shut}$. In \citet{2015A&A...578A..19H,2015ApJ...806..181H} we found that $\dot{M}_{\rm shut}=10\,M_{\rm Gan}\,{\rm Myr}^{-1}$ positions the H$_2$O ice line between Europa and Ganymede for a broad range of $\kappa_{\rm P}$ values.

To assess the effect of heating on the position of the H$_2$O ice line in the CPD as a function of solar distance, we compare two CSD models. First, we use the model of \citet[][H81, his Eq.~2.3]{1981PThPS..70...35H}, which assumes that the CSD is mostly transparent in the optical. Second, we use the model for an optically thick disk of \citet[][B15, their Eqs.~A.3 and A.7]{2015A&A...575A..28B}, which takes into account viscous and stellar heating as well as radiative cooling and opacity transitions. We consider both a low-metallicity ($Z=0.001$) and a solar-metallicity star ($Z=0.02$) with a stellar accretion rate $\dot{M}_\star=3.5\times10^{-8}\,M_\odot\,{\rm yr}^{-1}$ corresponding to Jupiter's runaway accretion phase \citep{2013A&A...558A.113M}, in which the planet opens up a gap and moon formation shuts down.

\begin{figure}[t]
   \centering
   \scalebox{.91}{\includegraphics{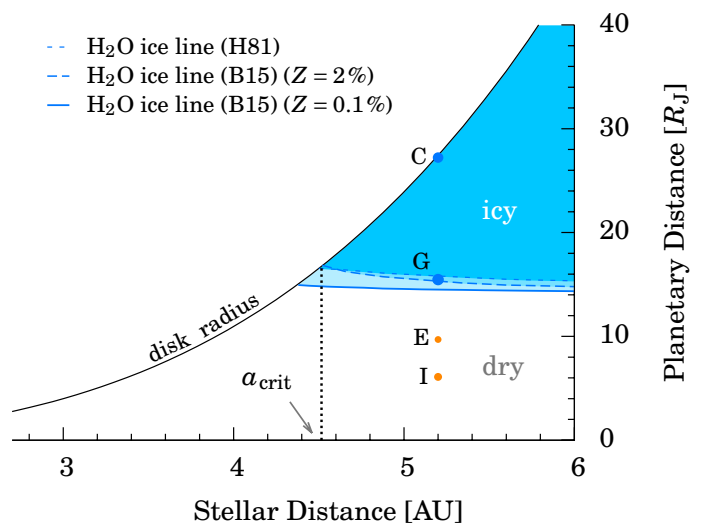}}
   \caption{Radial distance of Jupiter's H$_2$O ice line (roughly horizontal lines) and centrifugal disk radius (curved solid line) during its final accretion phase. Inside about 4.5\,AU from the Sun, Jupiter's disk does not contain an H$_2$O ice line in any of the CSD models \citep[][see legend]{1981PThPS..70...35H,2015A&A...575A..28B}. Locations of the Galilean moons are indicated by symbols at 5.2\,AU. Symbol sizes scale with the physical radii of the moons (orange: rocky, blue: icy composition).}
   \vspace{-.4cm}
   \label{fig:r_ice}
 \end{figure}

Figure~\ref{fig:r_ice} shows the results of our calculations, assuming fiducial values $\kappa_{\rm P}=10^{-2}\,{\rm m}^{-2}\,{\rm kg}^{-1}$ and $\dot{M}_{\rm shut}=10\,M_{\rm Gan}\,{\rm Myr}^{-1}$ as well as a centrifugal CPD radius as per \citet{2008ApJ...685.1220M}. The abscissa denotes distance from the young Sun, the ordinate indicates distance from Jupiter. The vertical dotted line highlights the critical solar distance ($a_{\rm crit}$) at which the CPD loses its H$_2$O ice line in the $Z=2\,\%$ H15 model. At 5.2\,AU, Callisto's position (circle labeled ``C'') at the outer CPD edge is due to our specific CPD scaling \citep{2015ApJ...806..181H}, whereas Ganymede's position (circle labeled ``G'') near the H$_2$O ice line is not a fit but a result.\footnote{\citet{2015ApJ...806..181H} argue that this suggests that Jupiter's H$_2$O ice line acted as a moon migration trap for Ganymede.} The CPD radius shrinks substantially towards the Sun owing to the increasing solar gravitational force, while the ice line recedes from the planet as a result of enhanced stellar heating. A comparison of the different CSD models (see legend) indicates that variations of stellar metallicity or solar illumination have significant effects on the circumjovian ice line, but the critical effect is the small CPD radius in the solar vicinity.

Most importantly, Fig.~\ref{fig:r_ice} suggests that Ganymede and Callisto cannot have accreted their icy components as long as Jupiter was closer than about $a_{\rm crit}=4.5$\,AU to the faint Sun, where the H$_2$O ice line around Jupiter vanishes for both the H81 and the B15 model. We varied $\kappa_{\rm P}$ and $\dot{M}_{\rm shut}$ by an order of magnitude (not shown), which resulted in changes of this critical solar distance of $\lesssim0.5$\,AU. Moreover, closer than 5.2\,AU from the Sun, Callisto's contemporary orbit around Jupiter would have been beyond the CPD radius, though still within Jupiter's Hill sphere.

Hence, large parts of both Ganymede and Callisto cannot have formed during the GT, which supposedly brought Jupiter much closer to the Sun than $4.5\pm0.5$\,AU. Thereafter, Jupiter's accretion disk (if still present at that time) would have been void of H$_2$O because water first would have vaporized and then been photodissociated into hydrogen and oxygen. If Ganymede or Callisto had acquired their H$_2$O from newly accreted planetesimals after the GT \citep[e.g. through gas drag within the CDP,][]{2003Icar..163..198M}, then Io (at 0.008 Hill radii, $\,R_{\rm H}$) and Europa (at $0.01\,R_{\rm H}$) would be water-rich, too, because planetesimal capture would have been efficient between 0.005 and $0.01\,R_{\rm H}$ \citep{2014ApJ...784..109T}. Hence, Ganymede and Callisto must have formed prior to the GT, at least to a large extent.

\vspace{-.15cm}

\subsection{Atmospheric escape from the Galilean moons}
\label{sub:escape}

Not only must a successful version of the GT model explain the observed H$_2$O ice contents in the Galilean moons, it also must be compatible with both the observed absence of thick atmospheres on the Galilean moons and the presence of a massive N$_2$ atmosphere around Saturn's moon Titan.

Several effects can drive atmospheric escape: (1) thermal (or ``Jeans'') escape, (2) direct absorption of high-energy (X-ray and ultraviolet, XUV) photons in the upper atmosphere, (3) direct absorption of high-energy particles from the solar wind, (4) impacts of large objects, and (5) drag of heavier gaseous components (such as carbon, oxygen, or nitrogen) by escaping lighter constituents (such as hydrogen) \citep{1987Icar...69..532H,2010ppc..book.....P}.\footnote{Atmospheric drag (5) can only occur if any of the other effects (1)--(4) is efficient for a lighter gas component.} Although additional chemical and weather-related effects can erode certain molecular species \citep{2006P&SS...54.1177A}, we identify the XUV-driven non-thermal escape during the GT as a novel picture that explains both the absence and presence of Galilean and Titanian atmospheres, respectively.

 \begin{figure}[t]
   \centering
    \scalebox{0.9}{\includegraphics{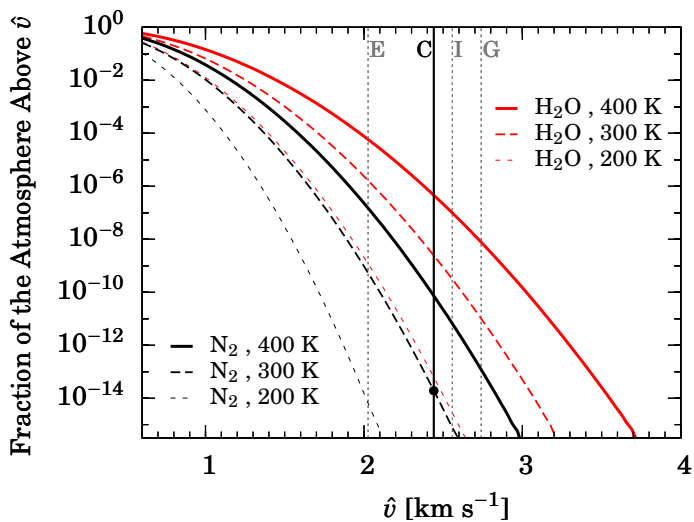}}
   \caption{Integrated Maxwell--Boltzmann velocity distribution of N$_2$ and H$_2$O gas molecules with temperatures similar to Callisto's surface temperature during accretion. The black vertical line denotes Callisto's escape velocity, values for the other Galilean moons are indicated by gray vertical lines. The black circle refers to an example discussed in the text.}
   \vspace{-.3cm}   
   \label{fig:v_th}
 \end{figure}

\subsubsection{Thermal escape}

As an example, we assess Callisto's potential to lose an initial N$_2$ or H$_2$O vapor atmosphere via thermal escape. Our choice is motivated by Callisto's relatively low surface temperature during accretion, which was about 300\,K at most, taking into account heating within Jupiter's accretion disk and from the accretion of planetesimals \citep{1982Icar...52...14L}. The other moons might have been subject to substantial illumination by the young Jupiter, so the effect of thermal escape might be harder to assess.

We compute $F_{\rm MB}(\hat{v})=\int_{\hat{v}}^\infty dv f_{\rm MB}(v)$, with $f_{\rm MB}(v)$ as the Maxwell--Boltzmann velocity distribution, for various gas temperatures $v$. Then $F_{\rm MB}^{\rm norm}(\hat{v}){\equiv}F_{\rm MB}(\hat{v})/F_{\rm MB}(0)$ is an approximation for the fraction of the atmosphere that is above a velocity $\hat{v}$.
 
In Figure~\ref{fig:v_th} we compare $F_{\rm MB}^{\rm norm}(\hat{v})$ for N$_2$ (black lines) and H$_2$O (red lines) molecules at temperatures between 200\,K and 400\,K (see legend) with Callisto's gravitational escape velocity ($2.44\,{\rm km\,s}^{-1}$, black vertical line). As an example, $F_{\rm MB}^{\rm norm}(\hat{v}>2.44\,{\rm km\,s}^{-1})\lesssim10^{-14}$ for an N$_2$ troposphere at 300\,K (see black circle), suggesting that a negligible fraction of the atmosphere would be beyond escape velocity.

Taking into account collisions of gas particles, their finite free path lengths, and the extent and temperature of the exosphere, \citet{2010ppc..book.....P} estimates the loss time of N$_2$ from Titan as $\approx10^{24}$\,yr. For Callisto, which has about $80\,\%$ of Titan's mass, this timescale would only be reduced by a factor of a few \citep[][Eq.~8.37]{2010ppc..book.....P} even if it had been as close as 1.5\,AU from the Sun. At that distance, Callisto would have received about 40 times higher irradiation than Titan receives today (assuming an optically thin CSD), and its exosphere temperatures might have been about $40^{1/4}\approx2.5$ times higher than Titan's.

Hence, thermal escape cannot be the reason for the lack of an N$_2$ atmosphere on Callisto, even during the GT as close as 1.5\,AU from the Sun.

\vspace{-.1cm}

\subsubsection{XUV-driven non-thermal escape and drag}

The X-ray and UV luminosities of the young Sun were as high as $10^2$ \citep{2005ApJ...622..680R} and $10^4$ \citep{1982RvGSP..20..280Z} times their current values, respectively, raising the question whether direct absorption of high-energy photons or atmospheric drag would have acted as efficient removal processes of an early Callistonian  atmosphere. Such a non-thermal escape would have eroded a hypothetical initial N$_2$ atmosphere from Earth in only a few $10^6$\,yr \citep{2010Icar..210....1L}, owing to the high exobase temperatures (7000-8000\,K) and the significant expansion of the thermosphere above the magnetopause \citep{2008JGRE..113.5008T}. However, the Earth's primitive atmosphere was likely CO$_2$-rich, which cooled the thermosphere and limited the N$_2$ outflow.

If Callisto initially had a substantial CO$_2$ or H$_2$O steam atmosphere, perhaps provided by outgassing, both gases would have been photo-dissociated in the upper atmosphere, which then would have been dominated by escaping H atoms; N$_2$ and other gases would have been dragged beyond the outer atmosphere and lost from the moon forever. \citet{2014OLEB...44..239L} simulated this gas drag for exomoons at 1\,AU from young Sun-like stars and found that the H, O, and C inventories in the initially thick CO$_2$ and H$_2$O atmosphere around moons 10\,\% the mass of the Earth (four times Ganymede's mass) would be lost within a few $10^5$\,yr depending on the initial conditions and details of the XUV irradiation. Increasing the moon mass by a factor of five in their computations, non-thermal escape times increased by a factor of several tens. Given that strong a dependence of XUV-driven escape on a moon's mass, Ganymede and Callisto would most certainly have lost initial N$_2$ atmospheres during the GT (perhaps within $10^4$\,yr), even if they approached the Sun as close as 1.5\,AU rather than 1\,AU as in \citet{2014OLEB...44..239L}.

\subsection{Titan's atmosphere in the Grand Tack model}
\label{sub:Titan}

If XUV-driven atmospheric loss from the Galilean moons indeed occurred during the GT, this raises the question why Titan is still surrounded by a thick N$_2$ envelope, since Saturn supposedly migrated as close as 2\,AU to the young Sun \citep{2011Natur.475..206W}. We propose that the key lies in the different formation timescales of the Galilean moons and Titan.

In the GT simulations of \citet{2011Natur.475..206W}, Saturn accretes about 10\,\% of its final mass over several $10^5$\,yr after its tack, when Jupiter is already fully formed. Hence, while the Galilean moons must have formed before Jupiter's GT (Sect.~\ref{sub:formation}) and migrated towards the Sun, thereby losing any primordial atmospheres, Titan formed after Saturn's tack and on a longer timescale. Titan actually must have formed several $10^5$\,yr after Saturn's tack, or it would have plunged into Saturn via its own type I migration within the massive CPD \citep{2006Natur.441..834C,2010ApJ...714.1052S} because there would be no ice line to trap it around Saturn \citep{2015ApJ...806..181H}.

The absence of N$_2$ atmospheres around the Galilean satellites supports the GT scenario and is compatible with the presence of a thick N$_2$ atmosphere around Titan. The Galilean satellites likely lost their primordial envelopes while approaching the Sun as close as 1.5\,AU, whereas Titan formed after Saturn's tack at about 7\,AU under less energetic XUV conditions. Its N$_2$ atmosphere then built up through outgassing of NH$_3$ accreted from the protosolar nebula \citep{Mandt}.

\vspace{-.2cm}

\section{Discussion}

Although the GT paradigm is still controversial, our results provide additional support for it. We show that the outcome of atmosphere-free icy Galilean satellites and of a thick N$_2$ atmosphere around Titan is possible in the GT scenario. This demonstrates how moon formation can be used to constrain the migration and accretion history of planets -- in the solar system and beyond. The detection of massive exomoons around the observed exo-Jupiters at 1\,AU from Sun-like stars would indicate that those gas giants migrated from beyond about 4.5\,AU.

We tested two CSD models (H81 and B15) to study the effect of changing solar heating on the radial position of the H$_2$O ice line in Jupiter's CPD. Both models yield similar constraints on the critical solar distance ($a_{\rm crit}$) beyond which the icy Galilean satellites must have formed. We varied opacities, shutdown accretion rates, and stellar metallicities to estimate the dependence of our results on the uncertain properties of the protoplanetary and protosatellite disks and found that variations of $a_{\rm crit}$ are $\lesssim0.5$\,AU. Residual heat from the moons' accretion, radiogenic decay, and tidal heating might have provided additional heat sources, which we neglected. Hence, $a_{\rm crit}=4.5\pm0.5$\,AU must be considered a lower limit.

If the GT actually took place and our conclusions about the pre-GT formation of the Galilean moons are correct, their early thermal evolution needs to be readdressed. Radiogenic decay in the rocky components and residual heat from accretion have been considered the main heat sources that determined post-accretion internal differentiation \citep{1987Icar...69...91K}. The total heat flux of several TW in both Ganymede and Callisto \citep{1988Icar...76..437M} translates into a few tens of ${\rm mW\,m}^{-2}$ on the surface. Our results, however, suggest that sunlight might have been a significant external energy source. At 1.5\,AU from the Sun, illumination might have reached several ${\rm W\,m}^{-2}$, if the CSD was at least partly transparent in the optical. Near-surface temperatures might have been approximately 10\,K higher for $\gtrsim10^5$\,yr \citep[e.g. in Fig.~3a of][]{2004Icar..169..402N}. The GT might have significantly retarded the cooling of the Galilean satellites.

Alternatively, Ganymede and Callisto might have become H$_2$O-rich after the GT, maybe through ablation of newly accreted planetesimals \citep{2010Icar..207..448M}, but then a mechanism is required that either prevented Io and Europa from accreting significant amounts of icy planetesimals or that triggered the loss of accreted ice. Tidal heating and the release of large amounts of kinetic energy from giant impacts might account for that.

As H is an effective UV absorber \citep{2004ApJ...615..972G} CSD gas might have shielded early Galilean atmospheres quite effectively. The net UV blocking effect depends on the disk scale height, the gas density profile in Jupiter's gap, and the residual gas flow \citep{2014ApJ...782...88F}. Ultraviolet photons might have been scattered deep into the gap, maybe even via a back-heating effect from the dusty wall behind the gap \citep{2010ApJ...710L.167H}. This aspect of our theory needs deeper investigation and dedicated CSD simulations with Jupiter migrating to 1.5\,AU.

\section{Conclusions}

In this Letter we show that the Grand Tack model for the migration of Jupiter and Saturn, if valid, imposes important constraints on the formation of their massive icy moons, Ganymede, Callisto, and Titan: \textbf{(1)} Ganymede and Callisto (probably also Io and Europa) formed prior to the GT (Sect.~\ref{sub:formation}), \textbf{(2)} their formation took place beyond $4.5\pm0.5$\,AU from the Sun (Sect.~\ref{sub:formation}), \textbf{(3)} the Galilean moons would have lost any primordial atmospheres during the GT via non-thermal XUV-driven escape due to the active young Sun (Sect.~\ref{sub:escape}), and \textbf{(4)} Titan's thick N$_2$ atmosphere and constraints from moon migration in CPDs suggest that Titan formed after Saturn's tack (Sect.~\ref{sub:Titan}).

Detailed observations of the Galilean moons by ESA's upcoming JUpiter ICy moons Explorer (JUICE), scheduled for launch in 2022 and arrival at Jupiter in 2030 \citep{2013P&SS...78....1G}, could deliver fundamentally new insights into the migration history of the giant planets. If Ganymede and Callisto formed prior to Jupiter's Grand Tack, then JUICE might have the capabilities to detect features imprinted during the moons' journey through the inner solar system.

\begin{acknowledgements}
We thank C.~Mordasini, B.~Bitsch, J.~Blum, D.~N.~C~Lin, W. Brandner, J.~Leconte, C.~P.~Dullemond, E.~Gaidos, and an anonymous referee for their helpful comments. RH is supported by the Origins Institute at McMaster U. and by the Canadian Astrobiology Program, a Collaborative Research and Training Experience Program by the Natural Sciences and Engineering Research Council of Canada (NSERC). REP is supported by an NSERC Discovery Grant.
\vspace{-.9cm}
\end{acknowledgements}


\bibliographystyle{aa} 
\bibliography{ms}




\end{document}